

\message
{JNL.TEX version 0.9 as of 3/27/86.  Report bugs and problems to Doug Eardley.}



\font\twelverm=cmr10 scaled 1200    \font\twelvei=cmmi10 scaled 1200
\font\twelvesy=cmsy10 scaled 1200   \font\twelveex=cmex10 scaled 1200
\font\twelvebf=cmbx10 scaled 1200   \font\twelvesl=cmsl10 scaled 1200
\font\twelvett=cmtt10 scaled 1200   \font\twelveit=cmti10 scaled 1200

\skewchar\twelvei='177   \skewchar\twelvesy='60


\def\twelvepoint{\normalbaselineskip=12.4pt plus 0.1pt minus 0.1pt
  \abovedisplayskip 12.4pt plus 3pt minus 9pt
  \belowdisplayskip 12.4pt plus 3pt minus 9pt
  \abovedisplayshortskip 0pt plus 3pt
  \belowdisplayshortskip 7.2pt plus 3pt minus 4pt
  \smallskipamount=3.6pt plus1.2pt minus1.2pt
  \medskipamount=7.2pt plus2.4pt minus2.4pt
  \bigskipamount=14.4pt plus4.8pt minus4.8pt
  \def\rm{\fam0\twelverm}          \def\it{\fam\itfam\twelveit}%
  \def\sl{\fam\slfam\twelvesl}     \def\bf{\fam\bffam\twelvebf}%
  \def\mit{\fam 1}                 \def\cal{\fam 2}%
  \def\tt{\twelvett}
  \textfont0=\twelverm   \scriptfont0=\tenrm   \scriptscriptfont0=\sevenrm
  \textfont1=\twelvei    \scriptfont1=\teni    \scriptscriptfont1=\seveni
  \textfont2=\twelvesy   \scriptfont2=\tensy   \scriptscriptfont2=\sevensy
  \textfont3=\twelveex   \scriptfont3=\twelveex  \scriptscriptfont3=\twelveex
  \textfont\itfam=\twelveit
  \textfont\slfam=\twelvesl
  \textfont\bffam=\twelvebf \scriptfont\bffam=\tenbf
  \scriptscriptfont\bffam=\sevenbf
  \normalbaselines\rm}



\def\beginlinemode{\endmode
  \begingroup\parskip=0pt \obeylines\def\\{\par}\def\endmode{\par\endgroup}}
\def\beginparmode{\endmode
  \begingroup \def\endmode{\par\endgroup}}
\let\endmode=\par
{\obeylines\gdef\
{}}
\def\singlespace{\baselineskip=\normalbaselineskip}

\def\oneandahalfspace{\baselineskip=\normalbaselineskip
  \multiply\baselineskip by 3 \divide\baselineskip by 2}
\def\doublespace{\baselineskip=\normalbaselineskip \multiply\baselineskip by 2}

\newcount\firstpageno
\firstpageno=2
\footline={\ifnum\pageno<\firstpageno{\hfil}\else{\hfil\twelverm\folio\hfil}\fi}
\def\toppageno{\global\footline={\hfil}\global\headline
  ={\ifnum\pageno<\firstpageno{\hfil}\else{\hfil\twelverm\folio\hfil}\fi}}
\let\rawfootnote=\footnote		
\def\footnote#1#2{{\rm\singlespace\parindent=0pt\parskip=0pt
  \rawfootnote{#1}{#2\hfill\vrule height 0pt depth 6pt width 0pt}}}
\def\raggedcenter{\leftskip=4em plus 12em \rightskip=\leftskip
  \parindent=0pt \parfillskip=0pt \spaceskip=.3333em \xspaceskip=.5em
  \pretolerance=9999 \tolerance=9999
  \hyphenpenalty=9999 \exhyphenpenalty=9999 }
\def\dateline{\rightline{\ifcase\month\or
  January\or February\or March\or April\or May\or June\or
  July\or August\or September\or October\or November\or December\fi
  \space\number\year}}
\def\received{\vskip 3pt plus 0.2fill
 \centerline{\sl (Received\space\ifcase\month\or
  January\or February\or March\or April\or May\or June\or
  July\or August\or September\or October\or November\or December\fi
  \qquad, \number\year)}}


\hsize=6.5truein
\vsize=8.9truein
\parskip=\medskipamount
\def\\{\cr}
\twelvepoint		
\doublespace		
\overfullrule=0pt	




\def\title			
  {\null\vskip 3pt plus 0.2fill
   \beginlinemode \doublespace \raggedcenter \bf}

\def\author			
  {\vskip 3pt plus 0.2fill \beginlinemode
   \singlespace \raggedcenter}

\def\affil			
  {\vskip 3pt plus 0.1fill \beginlinemode
   \oneandahalfspace \raggedcenter \sl}

\def\abstract			
  {\vskip 3pt plus 0.3fill \beginparmode
   \oneandahalfspace ABSTRACT: }

\def\endtopmatter		
  {\endpage			
   \body}

\def\body			
  {\beginparmode}		

\def\head#1{			
  \goodbreak\vskip 0.5truein	
  {\immediate\write16{#1}
   \raggedcenter \uppercase{#1}\par}
   \nobreak\vskip 0.25truein\nobreak}

\def\beneathrel#1\under#2{\mathrel{\mathop{#2}\limits_{#1}}}

\def\refto#1{$^{#1}$}		

\def\references			
  {\head{References}		
   \beginparmode
   \frenchspacing \parindent=0pt \leftskip=1truecm
   \parskip=8pt plus 3pt \everypar{\hangindent=\parindent}}

\gdef\refis#1{\item{#1.\ }}			

\gdef\journal#1, #2, #3, 1#4#5#6{		
    {\sl #1~}{\bf #2}, #3 (1#4#5#6)}		

\def\prb{\journal Phys. Rev. B, }

\def\prl{\journal Phys. Rev. Lett., }

\def\endreferences{\body}

\def\figurecaptions		
  {\endpage
   \beginparmode
   \head{Figure Captions}
}

\def\endpage			
  {\vfill\eject}

\def\endpaper			
  {\endmode\vfill\supereject}


\def\heading				
  {\vskip 0.5truein plus 0.1truein	
   \beginparmode \def\\{\par} \parskip=0pt \singlespace \raggedcenter}

\def\subheading				
  {\vskip 0.25truein plus 0.1truein	
   \beginlinemode \singlespace \parskip=0pt \def\\{\par}\raggedcenter}

\def\tag#1$${\eqno(#1)$$}

\def\align#1$${\eqalign{#1}$$}

\def\aligntag#1$${\gdef\tag##1\\{&(##1)\cr}\eqalignno{#1\\}$$
  \gdef\tag##1$${\eqno(##1)$$}}

\def\overset#1\to#2{{\mathop{#2}^{#1}}}
\def\underset#1\to#2{{\mathop{#2}_{#1}}}


\def\ref#1{Ref.~#1}			
\def\Ref#1{Ref.~#1}			
\def\[#1]{[\cite{#1}]}
\def\cite#1{{#1}}
\def\(#1){(\call{#1})}
\def\call#1{{#1}}
\def\taghead#1{}
\def\frac#1#2{{#1 \over #2}}

\def\12{{1\over2}}

\def\sla{\raise.15ex\hbox{$/$}\kern-.57em}
\def\leaderfill{\leaders\hbox to 1em{\hss.\hss}\hfill}
\def\twiddle{\lower.9ex\rlap{$\kern-.1em\scriptstyle\sim$}}
\def\bigtwiddle{\lower1.ex\rlap{$\sim$}}
\def\gtwid{\mathrel{\raise.3ex\hbox{$>$\kern-.75em\lower1ex\hbox{$\sim$}}}}
\def\ltwid{\mathrel{\raise.3ex\hbox{$<$\kern-.75em\lower1ex\hbox{$\sim$}}}}
\def\square{\kern1pt\vbox{\hrule height 1.2pt\hbox{\vrule width 1.2pt\hskip 3pt
   \vbox{\vskip 6pt}\hskip 3pt\vrule width 0.6pt}\hrule height 0.6pt}\kern1pt}
\def\tdot#1{\mathord{\mathop{#1}\limits^{\kern2pt\ldots}}}

\def\pmb#1{\setbox0=\hbox{#1}%
  \kern-.025em\copy0\kern-\wd0
  \kern  .05em\copy0\kern-\wd0
  \kern-.025em\raise.0433em\box0 }

\def\3he{{$^3${\rm He}}}

\def\slD{\raise.15ex\hbox{$/$}\kern-.57em\hbox{$D$}}
\def\dsl{\raise.15ex\hbox{$/$}\kern-.57em\hbox{$\Delta$}}
\def\slp{{\raise.15ex\hbox{$/$}\kern-.57em\hbox{$\partial$}}}
\def\nsl{\raise.15ex\hbox{$/$}\kern-.57em\hbox{$\nabla$}}
\def\sla{\raise.15ex\hbox{$/$}\kern-.57em\hbox{$\rightarrow$}}
\def\slla{\raise.15ex\hbox{$/$}\kern-.57em\hbox{$\lambda$}}
\def\slb{\raise.15ex\hbox{$/$}\kern-.57em\hbox{$b$}}
\def\lnp{\raise.15ex\hbox{$/$}\kern-.57em\hbox{$p$}}
\def\lnk{\raise.15ex\hbox{$/$}\kern-.57em\hbox{$k$}}
\def\lnK{\raise.15ex\hbox{$/$}\kern-.57em\hbox{$K$}}
\def\lnq{\raise.15ex\hbox{$/$}\kern-.57em\hbox{$q$}}
\def\lnA{\raise.15ex\hbox{$/$}\kern-.57em\hbox{$A$}}
\def\lna{\raise.15ex\hbox{$/$}\kern-.57em\hbox{$a$}}
\def\lnB{\raise.15ex\hbox{$/$}\kern-.57em\hbox{$B$}}


\def\pmb#1{\setbox0=\hbox{$#1$}%
\kern-.025em\copy0\kern-\wd0
\kern.05em\copy0\kern-\wd0
\kern-.025em\raise.0433em\box0 }

\def\q2{{Q^2}}
\def\gtwid{\raise.3ex\hbox{$>$\kern-.75em\lower1ex\hbox{$\sim$}}}
\def\ltwid{\raise.3ex\hbox{$<$\kern-.75em\lower1ex\hbox{$\sim$}}}
\def\12{{1\over2}}
\def\part{\partial}

\def\low#1{\lower.5ex\hbox{${}_#1$}}

\def\psl{\raise.15ex\hbox{$/$}\kern-.57em\hbox{$\partial$}}
\def\partt{\raise.15ex\hbox{$\widetilde$}{\kern-.37em\hbox{$\partial$}}}

\def\topppageno1{\global\footline={\hfil}\global\headline
={\ifnum\pageno<\firstpageno{\hfil}\else{\hss\twelverm --\ \folio
\ --\hss}\fi}}

\def\toppageno2{\global\footline={\hfil}\global\headline
={\ifnum\pageno<\firstpageno{\hfil}\else{\rightline{\hfill\hfill
\twelverm \ \folio
\ \hss}}\fi}}

\catcode`@=11
\newcount\r@fcount \r@fcount=0
\newcount\r@fcurr
\immediate\newwrite\reffile
\newif\ifr@ffile\r@ffilefalse
\def\w@rnwrite#1{\ifr@ffile\immediate\write\reffile{#1}\fi\message{#1}}

\def\writer@f#1>>{}
\def\referencefile{
  \r@ffiletrue\immediate\openout\reffile=\jobname.ref%
  \def\writer@f##1>>{\ifr@ffile\immediate\write\reffile%
    {\noexpand\refis{##1} = \csname r@fnum##1\endcsname = %
     \expandafter\expandafter\expandafter\strip@t\expandafter%
     \meaning\csname r@ftext\csname r@fnum##1\endcsname\endcsname}\fi}%
  \def\strip@t##1>>{}}

\def\citeall#1{\xdef#1##1{#1{\noexpand\cite{##1}}}}
\def\cite#1{\each@rg\citer@nge{#1}}     

\def\each@rg#1#2{{\let\thecsname=#1\expandafter\first@rg#2,\end,}}
\def\first@rg#1,{\thecsname{#1}\apply@rg}       
\def\apply@rg#1,{\ifx\end#1\let\next=\relax
\else,\thecsname{#1}\let\next=\apply@rg\fi\next}

\def\citer@nge#1{\citedor@nge#1-\end-}  
\def\citer@ngeat#1\end-{#1}
\def\citedor@nge#1-#2-{\ifx\end#2\r@featspace#1 
  \else\citel@@p{#1}{#2}\citer@ngeat\fi}        
\def\citel@@p#1#2{\ifnum#1>#2{\errmessage{Reference range #1-#2\space is bad.}%
    \errhelp{If you cite a series of references by the notation M-N, then M and
    N must be integers, and N must be greater than or equal to M.}}\else%
 {\count0=#1\count1=#2\advance\count1
by1\relax\expandafter\r@fcite\the\count0,%

  \loop\advance\count0 by1\relax
    \ifnum\count0<\count1,\expandafter\r@fcite\the\count0,%
  \repeat}\fi}

\def\r@featspace#1#2 {\r@fcite#1#2,}    
\def\r@fcite#1,{\ifuncit@d{#1}
    \newr@f{#1}%
    \expandafter\gdef\csname r@ftext\number\r@fcount\endcsname%
                     {\message{Reference #1 to be supplied.}%
                      \writer@f#1>>#1 to be supplied.\par}%
 \fi%
 \csname r@fnum#1\endcsname}
\def\ifuncit@d#1{\expandafter\ifx\csname r@fnum#1\endcsname\relax}%
\def\newr@f#1{\global\advance\r@fcount by1%
    \expandafter\xdef\csname r@fnum#1\endcsname{\number\r@fcount}}

\let\r@fis=\refis                       
\def\refis#1#2#3\par{\ifuncit@d{#1}
   \newr@f{#1}%
   \w@rnwrite{Reference #1=\number\r@fcount\space is not cited up to now.}\fi%
  \expandafter\gdef\csname r@ftext\csname r@fnum#1\endcsname\endcsname%
  {\writer@f#1>>#2#3\par}}

\def\ignoreuncited{
   \def\refis##1##2##3\par{\ifuncit@d{##1}%
     \else\expandafter\gdef\csname r@ftext\csname
r@fnum##1\endcsname\endcsname%

     {\writer@f##1>>##2##3\par}\fi}}

\def\r@ferr{\endreferences\errmessage{I was expecting to see
\noexpand\endreferences before now;  I have inserted it here.}}
\let\r@ferences=\references
\def\references{\r@ferences\def\endmode{\r@ferr\par\endgroup}}

\let\endr@ferences=\endreferences
\def\endreferences{\r@fcurr=0
  {\loop\ifnum\r@fcurr<\r@fcount
    \advance\r@fcurr by 1\relax\expandafter\r@fis\expandafter{\number\r@fcurr}%
    \csname r@ftext\number\r@fcurr\endcsname%
  \repeat}\gdef\r@ferr{}\endr@ferences}


\let\r@fend=\endpaper\gdef\endpaper{\ifr@ffile
\immediate\write16{Cross References written on []\jobname.REF.}\fi\r@fend}

\catcode`@=12

\citeall\refto          
\citeall\ref            %
\citeall\Ref            %

\def \q {\Psi^{(2)}([z_i],[\eta_{\alpha}])}

\def \ps1 {\psi^{\dag}_{i \alpha}}
\def \ps  {\psi_{j \beta}}

\def \fs  {``fractional statistics" \  }
\title Quaternion Generalization of the Laughlin State and the Three
Dimensional Fractional QHE.

\author A. V.  Balatsky$^{\dag}$
\affil Department of Physics
       University of Illinois at Urbana-Champaign,
       1110 W.Green St., Urbana IL 61801
\centerline{and}
\affil Landau Institute for Theoretical Physics, Moscow, USSR.

\abstract{ The 3D state of strongly correlated electrons is proposed,
which in the external magnetic field $\vec B$ exhibits the fractional
quantum Hall effect, with the zero temperature conductivity tensor
$\sigma_{ij} = (e^2/h)(1/m) \sum_k \epsilon_{ijk} B^k/\mid \vec B\mid
$. The analog of Landau and Laughlin states in 3D are given using
quaternion coordinates as generalization of complex coordinates. We
discuss the notion of the fractional statistics in 3D introduced
recently by Haldane.}

\body
\vskip .15in
\noindent PACS No. 05.-30.- d; 72.20.- i.
\endtopmatter

The concept of ``anyons" or ``fractional statistics" \refto{Anyon} particles
has been intensively  investigated recently. Soon after
experimental discovery of the Fractional QHE (FQHE)\refto{Exp} Laughlin
proposed variational wave function which describes the incompressible electron
liquid in 2D in external magnetic field at fractional filling factors,
which naturally leads to the \fs of the quasiparticles \refto{Laugh}. The  {\it
holomorphic}  structure of the  Laughlin state  rely essentially on the
fact that the coordinate space of electron liquid is 2D.
Mathematically 2D space allows the existence of the abelian representations
of the braid group instead of permutation group in higher dimensions \refto{
Braid}
.  Physically   Wilczek's construction of anyons as the  bound state of
particles with fluxes is very natural namely in 2D space.

However we were led to a conclusion that some aspects of the 2D FQHE state
can be generalized on to strongly correlated states of electrons in higher
dimensions, namely the 3D.  I will present the variational wave function
of the isotropic 3D electron liquid subjected in to strong magnetic field
$\vec B$.  This state can be considered as a generalization of
the Laughlin state on 3D electron liquid.  The conductivity tensor of this
state generalizes the FQHE conductivity:
$$\sigma_{ij} = {e^2 \over h} \  {1\over m} \ \sum_k \epsilon_{ijk} e^k
\eqno(1)
$$
with m-odd integer, and $\epsilon_{ijk}$ - antisymmetric tensor.  Analogous
formula
for IQHE in 3D has been obtained by Halperin\refto{Halp}
and Kunz\refto{Kunz}.  Detailed numerical calculations for noninteracting
particles were done by Montambaux
and Kohmoto \refto{MK}.
{}.

This state supports the fractional statistics of the quasiparticle
excitations, which have nontrivial braiding phases.  Presence of external
magnetic field $\vec B$ provides the external axis in otherwise
isotropic 3D space what allows the braid group classification of the
particles paths in the plane $\pi_{\vec B}$, perpendicular to
the magnetic field $\vec B$.

  I will also consider the
vortex-like quasiparticle wave function in 3D which is a candidate for
fractional
statistics in Haldane's definition \refto{ Stat}
{}.

{\bf Generalization of the Landau state and IQHE.}

Consider first the generalization of the lowest Landau level (LLL) wave
function on 3D space.  In 2D case the Landau wave function is written in
terms of {\it holomorphic} coordinates $z_j = x_j + i y_j, j = 1, . . .N$ is
the particle number.  The LLL wave function
$$ \Psi^{2D}_{LLL} = \prod_{i<j} (z_i - z_j) e^{-{1\over 4}\mid z_i
\mid^2} \eqno(2)$$
satisfies the following conditions:

\item{(a)}it is  odd under any permutations;
\item{(b)}it is  isotropic;
\item{(c)}it  has a Jastrow form and is homogenous polynomial.

\noindent
To construct the analog of Landau state in 3D space, now consider coordinates
$x_j, y_j, z_j$ in 3D.  (There should be no confusion between $z_i$ - the
third coordinate in 3D space and complex 2D coordinate $z_j = x_j + iy_j$.
To generalize 2D {\it holomorphic} coordinates on 3D space, introduce complex
quaternions $q \in {\cal H}$.
$$
q = \hat i x + \hat j y + \hat k z, \bar q = - q \eqno (3a)
$$

$$
\hat i \hat j = -\hat j \hat i = \hat k, \hat j \hat k = \hat i = -\hat k
\hat j, \hat k \hat i = \hat j = \hat {-i} \hat k
$$

$$
\hat {i^2} = \hat {j^2} = \hat {k^2} = -1\eqno(3b)
$$
Complex quaternions realize the mapping of configuration of the particles in 3D
space onto the ${\cal H}$ space of quaternions.  For each position of  spinless
particles
$x_i, y_i, z_i, i = 1, . . .N$, consider corresponding weight with $q_i =
\hat i x_i + \hat j y_i + \hat k z_i$

$$
\Psi^{3D}_{LLL} (r_i) = : \prod_{i<j} (q_i - q_j):\eqno(4)
$$
Where because of noncommutativity of multiplication in ${\cal H}$, the
ordering of multiplication is introduced.  For example, one may choose to
multiply $q_1$ with all $q_i, i>1$ first, then $q_2$ with $q_i, i>2$, {\it
etc.} in the ordered way in (4).  The ``wave function" $\Psi^{3D}_{LLL}$ is a
3D generalization of the Landau state (its {\it holomorphic} part more
precisely). It obviously satisfies the conditions a) - c) mentioned above.
However it has no physical meaning itself because
$\Psi^{3D}_{LLL}$ is $SU(2)$ valued. We will project out extra degrees of
freedom in
it using the fact that external magnetic field provides natural coordinate
axis. This operation corresponds to  taking $U(1)$  subgroup of $SU(2)$.

It is  convenient to use the isomorphism of quaternion algebra ${\cal H}$ -
to the algebra of Pauli matricies:
$$
\hat i = -i \sigma^1; \hat z = -i \sigma^2; \hat k = -i \sigma^3 \eqno(5)
$$
Then $\Psi^{3D}_{LLL}$ can be rewritten as
$$
\Psi^{3D}_{LLL} (r_i) =: \prod_{i<j} \mid \vec r_i - \vec r_j\mid e^{-i {\vec
\sigma} \cdot {\vec h}_{ij} \pi/2}: exp (-{1 \over 4} \mid
q_i\mid^2) \eqno(6)
$$

$$
=:\prod_{i<j} (\sigma^1 x_{ij} + \sigma^2 y_{ij} + \sigma^3 z_{ij}): exp
(-{1\over 4} \mid q_i \mid^2)
$$
with ${\vec h}_{ij} = {{\vec r}_i - {\vec r}_j \over \mid {\vec r}_i - {\vec
r}_j \mid}
, \  x_{ij} = x_i = x_j, \  y_{ij} = y_i - y_j$ and $ z_{ij} = z_i - z_j$.

To make the plasma analogy, so useful in 2D, applicable, the $\Psi^{3D}_{LLL}
(r_i)$ is multiplied by a scalar factor $\prod_i exp (-{1\over 4} \mid q_i
\mid^2)$.  This will ensure the {\it incompressibility} of this state after
projection (see below) and
enables to construct vortex like quasiparticles.Eq.(6) can be generalized to $
\Psi^{3D} (
r_i) =: \prod_{i<j}f(
 \mid \vec r_i - \vec r_j\mid) e^{-i {\vec
\sigma} \cdot {\vec h}_{ij} \pi/2}: exp (-{1 \over 4} \mid
q_i\mid^2) $ with $f(x)$ being some general scalar function. For $f(x) =
e^{-1\over {x}}$ the plasma analogy can be applied to  3D state.
 Hamiltonian $H_{pl} = -{1\over {\beta}} ln\langle \Psi^{3D}(r_i)\mid
\Psi^{3D}(r_i)\rangle $ becomes a one-component 3D plasma Hamiltonian.

  Next crucial step is to
project the "wave function" $\Psi^{3D}_{LLL} (r_i)$ down on to space of
complex numbers, to construct wave function of 3D system in external magnetic
field.  So far $\Psi^{3D}_{LLL} (r_i)$ is matrix valued ( SU(2) essentially),
as follows from Eq.
(6)\refto{Matrix}
.  Introduce the projectors:
$$
P_z = \sigma^1 + i \sigma^2, \  Tr(P_z q_{ij}) = x_{ij} + i y_{ij}
$$

$$
P_x = \sigma^2 + i \sigma^3, \  Tr(P_x q_{ij}) = y_{ij} + z_{ij} \eqno(7)
$$

$$
P_y = \sigma^3 + i \sigma^1, \  Tr(P_y q_{ij}) = z_{ij} + i x_{ij}
$$
and define
$$
P_z \Psi^{3D}_{LLL} (r_i) \equiv :\prod_{i<j} (Tr P_z q_{ij}): e^{-{1 \over 4}
\mid
Tr P_z q_{i}\mid^2} \eqno (8)
$$

$$
= \Psi^{2D}_{LLL} (r_i)
$$

\noindent
Where in the l.h.s. $P_z$ is an operator, and its action on coordinates is
given by Eq(7). Below we will drop the sign of $Tr$ since it is obvious.
The role of these projector operators $P_{i} , i =  x, y, z$ is to
discriminate between three equivalent coordinate axis in the presence of
external magnetic field $\vec B$.  Physically the projected state
coincides with the Landau state $P_z \Psi^{3D}_{LLL} (r_i) =  \Psi^{2D}_{LLL}
(r_i)$, and describes the IQHE of the 3D isotropic sample of
correlated matter subjected to strong external magnetic field $\vec B \parallel
\hat Z$.  The effect of external field on the orbital
motion of the particles is assumed stronger then any internal correlations,
incorporated in $\Psi^{3D}_{LLL} (r_i)$.  For example the correlations along
$z$ direction are completely ignored in this projected state.

Because of {\it incompressibility} of  $\Psi^{2D}_{LLL}$ projected
state $P_z\Psi^{3D}_{LLL} (r_i)$  is also {\it incompressible} as it should be.
 It
means that chemical potential of the particles lies in the gap.  The Hall
conductance of projected state is nothing more than the  conductance of
the LLL:
$$
\sigma_{xy} = {e^2 \over h} \eqno (9)
$$

For any unit vector $\vec e = e_1 \hat x + e_2 \hat y + e_3 \hat
z$ define the projector $P_{\vec e}$:
$$
P_{\vec e} = e_1  P_x + e_2 P_y + e_3 P_z \eqno (10)
$$
Then the corresponding projected state $P_{\vec e}
\Psi^{3D}_{LLL} (r_i)$ is:
$$
P_{\vec e} \left (\Psi^{3D}_{LLL} (r_i)\right ) =
\sum_i e_i P_i \left (\Psi^{3D}_{LLL} (r_i)\right ) = \eqno (11)
$$

$$
= \sum_i e_i \left (\Psi^{2D}_{LLL} (r_i)  \  {for} \vec B
\parallel \hat i\right )
$$
{}From Eqs. (9) - (11) follows that in ${\vec B} = {\vec e} \mid {\vec B}
\mid$,
 conductivity tensor of the state
described by Eq. (11) is:
$$
\sigma_{ij} = {e^2 \over h} \ \sum_k \epsilon_{ijk} e_k \eqno (12)
$$
Eq. (12) is a generalization of the IQHE conductivity for 3D
electron liquid in magnetic field.

Closely related formula has been discussed previously \refto{Halp,Kunz, MK}
in the context of the conductivity of 3D electron gas in external periodical
potential.  Because of continuum limit taken from the beginning, instead of
reciprocal lattice vectors in the case of external potential,
 in Eq. (12) enters the unit vector of external
field.  It is important to realize that the gap in state $\Psi^{3D}_{LLL}
(r_i)$ and its fractional
generalizations (see below) is produced by internal electron correlations
, in contrast to gap due to external potential in
\refto{Halp, Kunz, MK}.

 {\bf Generalizations of the Laughlin state and 3D FQHE}.

 Generalization of the Laughlin state
$$
\Psi^{2D}_L (r_i) = \prod_{i<j} (z_i - z_j)^m exp (-{1\over 4} \mid z_i
\mid^2) \eqno(13)
$$
is not unique and can be done in two distinct ways.

Firstly, consider 3D Laughlin state in strong magnetic field ${\vec B}
\parallel {\vec e}$ as a projected state defined by:
$$
P_{\vec e} \Psi^{3D}_L (r_i) = \sum^3_{i=1} e_i P_i \Psi^{3D}_L (r_i) =
$$

$$
= \sum_i e_i: \prod_{k<\ell} \left (P_i (q_k - q_\ell)\right )^m: exp\left
(-{1 \over 4} \mid P_i  q_k\mid^2\right ) \eqno (14)
$$

For example, $P_x \Psi^{3D}_L (r_i) = \prod_{k<\ell} (y_{k\ell} +
iz_{k\ell})^m exp \left (-{1 \over 4} (  y^2_k + z^2_k)\right )$.
Projection on each axis leads to the $2D$ Laughlin state for electrons in
magnetic field, parallel to this axis.  Because of incompressibility of the
projected state $P_{\vec e} \Psi^{3D}_L (r_i)$ for any $\vec e$, the chemical
potential always lies in the gap induced by electron correlations in this 3D
state.  The conductivity tensor is given by Eq. (1).

 Quasihole wave function
is given by the sum of Laughlin's quasihole wave functions written in proper
coordinates for each projection:
$$
P_{\vec e} \Psi^{3D}_{q_o} (r_i) = \sum_i e_i: \left( \prod_k P_i (q_o -
q_k)\right ) \prod_{k<\ell} \left (P_i (q_e - q_k)\right )^m:   exp (-{1 \over
4} \mid P_i q_k\mid^2 ) \eqno (15)
$$
$q_o$ is a position of the quasihole in $3D$ space.  The vortex character of
quasihole and quasiparticle as well as their nontrivial braiding properties
naturally follows from the form of Eq. (15).  For example, from $P_x
\Psi_{q_o, q\prime_o}^{3D} (r_i) = \prod_k (y\prime_{ok} + iz\prime_{ok})
(y_{ok} + iz_{ok}).  \prod_{k<\ell} (y_{k\ell} + iz_{k\ell})^m exp(-{1\over 4}
\mid   y^2_k + z^2_k\mid)$, it follows that adiabatic exchange of two
quasiholes, located at $q_o$ and $q\prime_o$ in $3D$ brings the phase $\pi
\over m$ per quasihole if the exchange was done along the path having
nontrivial projection on the plane $(y,z)$ \refto{Nontriv}.  Indeed, despite in
$3D$ space, it is acceptable to classify the paths of the particles with
respect to their winding numbers.  The reason for this classification to be
well defined, and then leading to the ``fractional statistics" is the
existence of the magnetic field axis in $3D$ space what makes the relevant
geometry to be $2D$ after projection.  Clearly the same consideration can be
done for any direction of the magnetic field.

Although the projection of $\Psi^{3D}_L (r_i)$ is defined as Eq. (15), I can
not guess the form of $\Psi^{3D}_L (r_i)$ which leads to Eq. (15).  The
straightforward generalization of the Laughlin wave function gives the second
possibility to define $3D$ Laughlin state with  different physical
consequences.  Define $SU(2)$ (mod scalar function) valued ``weight" as:
$$
{\tilde \Psi}^{3D}_L (r_i) = :\prod_{i<j} (q_i - q_j)^m: exp (-{1\over 4}
\mid q_i\mid^2)\eqno (16)
$$
Because of $(q_i -q_j)^m  = \mid \vec r_i -\vec r_j\mid^{m-1} (q_i - q_j)$ this
state is
equivalent to
$$
{\tilde \Psi}^{3D}_L (r_i) = :\prod_{i<j} \mid \vec r_i - \vec r_j\mid^{m-1}
(q_i -
q_j) : exp (-{1 \over 4} \mid q_i\mid^2). \eqno (17)
$$
Projected state
$$
P_{\vec e} {\tilde \Psi}^{3D}_L (r_i) = \sum^3_{i=1} e_i : \prod_{k<\ell} \mid
P_i \vec r_{k\ell}
\mid^{m-1} \left (P_i (q_k - q_\ell) \right ) : exp (-{1\over 4} \mid P_i
q_k\mid^2) \eqno (18)
$$
has a form of IQHE state with modified orbital factor, however it does not
have a LLL wave function structure.  Probably this state can be obtained
using appropriately chosen Haldane's pseudopotentials \refto{Pseudopot} in $2D$
problem.  Thought the {\it holomorphic} part of the wave function has a simple
LLL structure, the density of particles in this state $\rho(r) =
<P_{\vec e}{\tilde \Psi}^{3D}_L (r_i)\mid \sum_i \delta (\vec r_i - \vec r)\mid
P_{\vec e}{\tilde \Psi}^{3D}_L (r_i)>$ is $\rho_{LLL}\over m $.  The quasihole
state
in ${\tilde \Psi}^{3D}_L (r_i)$ is given by
$$
{\tilde \Psi}^{3D}_{q_o} (r_i) = :\prod_{k<\ell} (q_o - q_k) (q_k - q_\ell)^m
: exp (-{1\over 4} \mid q_i \mid^2). \eqno (19)
$$
For the projected quasihole state $P_{\vec e} {\tilde \Psi}^{3D}_{q_o} (r_i)$
the Berry phase argument leads to the statistics $\pi/m$ for quasiparticles.
(See Appendix A)  This is a somewhat surprising result, since all our insight
comes from the Laughlin states, where each electron corresponds to the $m$
fluxes, while quasihole corresponds to unit flux what explains $1/m$ in
charge and statistics of quasiparticles in FQHE.  It turns out that the
crucial feature of the FQHE to support ``fractional statistics" is {\it
incompressibility} which fixes the particle density deficit at quasiparticle
positions to be $\pm  {1\over m}$.  The original particles does not have to
be bound to $m$ fluxes in order to get ``fractional statistics" of
quasiparticles.

As a possible example of fractional statistics in Haldane's definition
\refto{Stat}, consider the  simplest quasiparticle state in 3D space given by
$\tilde \Psi^{3D}_{q_o} (r_i)$.  Although this object is not a wave function,
it is not relevant for the discussion below.  Winding properties of
quasiparticles at positions $q_o$ and $q\prime_{o}$ are trivial
\refto{Unpublished} - the Berry phase for winding is given by the solid angle
$\gamma = \Omega_c$ subtended by a unit vector ${\vec n} = {{\vec r}_o -
{\vec r}_{o\prime} \over \mid {\vec r}_o - {\vec r}_{o\prime}\mid}$
transforming along contour $C$.  For any true permutation $\Omega_c = 2 \pi$,
what makes quasiparticles to be fermions, as well as underlying particles.
We just obtained the well known fact - in 3D the classification of the paths
of the particles is given by the permutation group $P_n$, which allows only
the fermi and bose statistics.  From this point of view fractional statistics
is forbidden in 3D.  However, recently Haldane proposed the generalization of
the notion of the fractional statistics, based on the counting number of
states in the Hilbert space for the quasiparticles\refto{Stat} which can be
relevant for this 3D state.

Consider quasiparticle excitation in the form, consistent with 2D FQHE
case\refto{Laugh}:

$$
\tilde \Psi^{3D}_{+q_o} (r_i) = :\prod_{k<\ell} (q_o - q_k)^{-1} \hat Q_{q_o}
(q_k - q_\ell)^m: exp (-{1\over 4} \mid q_i \mid^2) \eqno(20)
$$
with $q_o$ as the position of the quasiparticle and, $\hat Q_{q_o}$ as the
projection operator reviving states with $(q_k - q_o)^o exp (-{1\over 4} \mid
q_k \mid^2)$ from the many body state to avoid divergency in Eq.(20).  The
subscript + in $\tilde \Psi^{3D}_{+q_o}$ is introduced to distinguish between
quasiparticle state $\tilde \Psi^{3D}_{+q_o}$ and quasihole state $\tilde
\Psi^{3D}_{q_o} \equiv \tilde \Psi^{3D}_{-q_o}$.  Fixing boundary
conditions in 3D will lead to constraint
$$
Nm + N_+ - N_- = const \eqno(21)
$$
analogously to the 2D FQHE case\refto{Stat}.  Where $N$ is the number of
particle, $N_\pm$ is the number of quasiparticles and quasiholes in state
$\tilde\Psi^{3D} (r_i)$.  The constraint Eq.(21) survives after projection
operation on any axis, when $P_{\vec e} \tilde\Psi^{3D}_L (r_i)$ simply
becomes the Laughlin state.  If the dimension of the Hilbert space for
quasiparticles is given by $d=N$, changing the $N^{\pm}$ by multiples of $m$
will lead to the fractional statistics of quasiparticles in Haldane's form:
$$
\Delta d = \pm {1\over m} \Delta N^\pm \eqno(22)
$$

To summarize, the isotropic  3D state of correlated electron liquid in external
magnetic
field is proposed. This state exhibit the fractional Hall conductivity tensor
Eq.(1),
which is transversal to the external magnetic field at zero temperature.
Quite generally the form of the conductivity tensor given by Eq. (1) is fixed
by the isotropy of the original 3D state and the presence of the uniaxial
field $\vec B$.  The time reversal symmetry properties of $\sigma_{ij}$ and
$\vec B$ are the same - they both are odd  under $t\rightarrow -t$.
$\sigma_{ij}$ and $\vec B$ are also odd with respect to parity \underbar {P}
transformations. For a 3D electron system with the Fermi level lying in the
gap of the energy spectrum the conductivity tensor has to be antisymmetric,
as it is given by Eq. (1).  Moreover the components of the conductivity
tensor, corresponding to the electric field ${\vec E} \parallel {\vec B}$
does not enter since they cannot produce persistent currents and will be
screened
out.\refto{Projector}

 To
generalize the complex coordinates on to 3D the quaternion algebra is used.
Although the 3D weight is matrix valued, the projection on to plane
perpendicular to the external field is used to construct true complex wave
function which is the weighted sum of the Laughlin states in different planes.
The 3D analogs of the quasiparticle and quasihole excitations are found. Again,
presence of the magnetic field allows the braid group classification of the
particle paths what leads to ``fractional statistics". If  to consider
the unprojected state as ``wave function" for some problem, the winding
properties of the quasiparticles in 3D are trivial. However it is argued that
these excitations can have ``fractional statistics" in  Haldane's definition,
using counting of states in the Hilbert space\refto{Stat}.

Physically this kind of 3D  state  can be obtained, if to consider the set of
QHE planes, perpendicular, say to $\hat z$ \refto{MacDonald},
 with strong interplanar coupling which makes the problem essentially
3D. Then tilting of  the magnetic field may cause the coherent
 interplanar hopping of
electrons in order to build the cyclotron orbits in the external field. The
states constructed in this article can not be obtained perturbatively in
interplanar hopping matrix element from the Laughlin 2D state, and represent
the new phase of 3D electron liquid in external field.

\noindent
{\bf Acknowledgments}

I wish to thank S. Girvin for fruitful discussions, and A. H. MacDonald for
bringing the ref \refto{Halp, MacDonald} to my attention.  This work was
supported in
part by NSF grant DMR-8817613

\vfil\eject
\noindent
{\bf Appendix A}

We describe here the derivation of the statistics of the quasiparticles in
projected state $P_{\vec e} {\tilde \Psi}^{3D}_{q_o} (r_i)$, Eq. (18).
Choose ${\vec e} = {\hat z}$ for simplicity, the result will be independent on
orientation of the field, as long as contours involved have nonzero
projection onto the plane, perpendicular to the field.
$$
P_z {\tilde \Psi}^{3D}_{q_o} (r_i) = \prod_j (x_{oj} + i y_{oj}) \prod_{j<k}
(x_{jk} + i y_{jk}) \mid P_z \vec r_{jk}\mid^{m-1} \times exp (-{1\over 4}\mid
x^2_j +
y^2_j \mid ) \eqno(A.1)
$$
The Berry phase for adiabatic transport $q_o(t)$ along some contour $C$ in
$(xy)$ plane is
$$
\gamma = i \oint < P_z {\tilde \Psi}^{3D}_{q_o (t)} (r_i) \mid {d \over
{dt}}\mid P_z {\tilde \Psi}^{3D}_{q_o (t)} (r_i)> dt \eqno (A.2)
$$

The matrix element of time derivation in (A.2) can be rewritten as
$$
<P_z {\tilde \Psi}^{3D}_{q_o (t)} (r_i) \mid {d\over {dt}} \sum_j \ln
\left (x_{oj} (t) + i y_{oj} (t)\right ) \mid P_z {\tilde \Psi}^{3D}_{q_o (t)}
(r_i)>
$$

The Berry phase then is:
$$
\gamma = i \oint dz_o \int d^2 z  \ \rho (z) {1 \over z_o - z} = \pm {2\pi
\over m}
\rho_{LLL} \hbox {Area} \eqno (A.3)
$$
with $z = x + iy$, and using the fact that density of particles $\rho
(z) = {1\over m} \rho_{LLL}$ in state  $P_z {\tilde \Psi}^{3D}_L (r_i)$:

$$
\rho(z) = <P_z {\tilde \Psi}^{3D}_L (r_i) \mid \sum_i \delta (z-z_i) \mid P_z
{\tilde \Psi}^{3D}_L (r_i)>
$$

$$
= <\Psi^{2D}_L (r_i) \mid \sum_i \delta (z - z_i) \mid \Psi^{2D}_L (r_i)> =
\rho_{1/m}\eqno (A.4)
$$
And $\rho_{1\over m}$ is the density of the Laughlin state  Eq.(13).

\endpage

\references

$^{\dag}$  Address after 1 September 1991: Los Alamos National Laboratory,
Center for Material Studies, Los Alamos, NM 87545.

\refis{Projector} It is not crucial for the final answer whether the
Projector operator is in the exponential factor, such as in Eq. (14) or not.
All that it amounts to is the change of the shape of the electron liquid.

\refis{Anyon} J. M. Leinaas and J. Myrheim, {\it Nuovo Chimento}, {\bf 37B}, 1,
(1977).
 F. Wilczek, {\it Phys. Rev. Lett.}  {\bf 49}, 957, (1982).

\refis{Exp} D. C. Tsui, H. L. Stormer and A.C. Gossard, \prl 48, 1559, 1982.

\refis{Laugh} R. B. Laughlin, \prl 50, 1395, 1983.

\refis{Halp} B. I. Halperin, Proceedings of the LT-18, Kyoto, p.1913, (1987).

\refis{MK} G. Montambaux and M. Kohmoto, \prb 41, 11417, 1990.

\refis{Kunz} H. Kunz, \prl 57, 1095, 1986.

\refis{Unpublished} A. V. Balatsky, unpublished.

\refis{Stat} F.D.M. Haldane, Princeton preprint, 1991.

\refis{Matrix} $\Psi^{3D}_{LLL}(r_i)$ can be obtained probably in a matrix
quantum mechanics as a generalization of the Landau
state.

\refis{Nontriv} Nontrivial means that projected path is a loop with well
defined
direction, corresponding to the way quasiparticles are exchanged.

\refis{Pseudopot} F.D.M. Haldane,  in ``The Quantum Hall Effect", eds R. Prange
and S.
Girvin (Springer, Berlin, 1990) Chap.7.

\refis{Braid} See for example: Y.S. Wu, \prl 52, 2103, 1984; B. Block and
X.G.Wen, \prb
42, 8145, 1990; {\it ibid} p. 8133; G. Moore and N. Read to be published;
 A.
Balatsky and E. Fradki, \prb 43, 10622, 1991.

\refis{MacDonald} X. Qiu, R. Joynt and A. H. MacDonald, \prb 42, 1339, 1990.

\endreferences

\bye